\documentclass[amsmath,amssymb,prc,final,showpacs,twocolumn]{revtex4-1}

\usepackage{bm,hyphenat,xspace}
\usepackage{graphicx,epsfig}
\usepackage{color}

\newcommand{\fm}{\;\mathrm{fm}}
\newcommand{\cm}{\mathrm{c\!\:\!.m\!\:\!.}}
\newcommand{\A}[2]{{}^{#1}\mathrm{#2}}

\begin{document}

\title {
Weak sensitivity of three-body ($d,p$) reactions to $np$ force models
}

\author{A.~Deltuva}
\email{arnoldas.deltuva@tfai.vu.lt}
\affiliation
{Institute of Theoretical Physics and Astronomy,
Vilnius University, Saul\.etekio al. 3, LT-10257 Vilnius, Lithuania
}

\date{3 May, 2018}

\begin{abstract}
\begin{description}
\item[Background]
Adiabatic distorted-wave approximation (ADWA)
study of three-body $(d,p)$ transfer reactions 
[G.W. Bailey, N.K. Timofeyuk,
and J.A. Tostevin, Phys. Rev. Lett. 117, 162502 (2016)]
reported strong sensitivity of cross sections
to the neutron-proton $(np)$ interaction model
when the  nucleon-nucleus optical potential is nonlocal.
\item[Purpose]
The verification of this unusual finding  using 
 more reliable methods is aimed for in the present work.
\item[Methods]
A rigorous Faddeev-type three-body scattering 
theory is applied to the study of $(d,p)$ transfer reactions.
The equations for transition operators are solved in the
momentum-space partial-wave framework.
\item[Results]
Differential cross sections for
 $\A{26}{Al}(d,p)\A{27}{Al}$ reactions are calculated
using nonlocal nuclear optical potentials
and a number of realistic $np$ potentials.
Only a weak dependence on the $np$ force
model is observed, typically one order of magnitude lower
than in the previous ADWA study.
The shape of the angular distribution of the experimental data
is well reproduced.
\item[Conclusions]
Cross sections of $(d,p)$ transfer reactions calculated using
a rigorous three-body method show little sensitivity to
the $np$ interaction model. This indicates a failure of the ADWA
in the context of nonlocal  potentials. Some evident shortcomings
of the ADWA are pointed out.
\end{description}
\end{abstract}

 \maketitle

\section{Introduction}

Three-body $(d,p)$ transfer reactions 
with nucleon-nucleus $(NA)$ nonlocal optical potentials (NLOP) 
and a number of neutron-proton $(np)$ potentials have been
studied in Ref.~\cite{PhysRevLett.117.162502}
using  the adiabatic distorted-wave approximation (ADWA).
The authors of Ref.~\cite{PhysRevLett.117.162502}
 claimed that NLOP leads to a
substantial increase of the  sensitivity of low-energy ($d,p$)
 cross sections to high-momentum components in the 
deuteron wave function and  $np$ interaction. This sensitivity 
manifested itself as a strong dependence of $(d,p)$  cross sections
on the  $np$ potential model, even exceeding a factor of 2
in particular cases such as the
 $\A{26}{Al}(d,p)\A{27}{Al}$(g.s.) reaction.
Such sensitivity  contrasts with the usual behavior
in the low-energy nuclear physics and calls for further studies
using alternative methods. 
The authors of Ref.~\cite{PhysRevLett.117.162502}
suggested that the Weinberg state expansion, the basis of ADWA,
should be revisited in the presence of nonlocality. 
The present work, being an extended version
of the pioneering study \cite{deltuva:commdp}, goes well beyond the 
suggestion of Ref.~\cite{PhysRevLett.117.162502}.
I use a rigorous treatment of three-body  reactions with NLOP
based on the Faddeev theory \cite{faddeev:60a}.
Although much more complicated than ADWA
in terms of practical implementation \cite{deltuva:09b}, 
this treatment has the advantage of providing 
an exact solution of the three-body scattering problem.
It does not rely on the Weinberg state expansion, thus,
implicitly includes all Weinberg states up to infinity.
Furthermore, it includes also the so called "remnant term", 
neglected in ADWA of  Ref.~\cite{PhysRevLett.117.162502},
all relevant $np$ waves beside the deuteron wave,
and the proper Coulomb force acting between the proton and nucleus, 
not between the center-of-mass (c.m.) of the deuteron and nucleus 
as in the initial channel treatment by ADWA.

Section II recalls the three-body Faddeev formalism for transition
operators. The results for $\A{26}{Al}(d,p)\A{27}{Al}$ reactions,
the study of the $np$ force model sensitivity, and the comparison
with the experimental data are presented in Sec. III.
Section IV contains the conclusions and discussion, also pointing out 
some inadequacies of the ADWA.

\section{Theory}

A rigorous three-body treatment of nuclear reactions with NLOP
based on the Faddeev equations in the Alt-Grassberger-Sandhas (AGS)
form \cite{alt:67a} was first implemented in 
Refs.~\cite{deltuva:09b,deltuva:09d}.
At the given three-body  energy $E$ in the c.m. frame 
the transition operators
$U_{\beta \alpha}(E)$ obey the system of integral equations
\begin{equation}  \label{eq:Uba}
U_{\beta \alpha}(E)  = \bar{\delta}_{\beta\alpha} \, G^{-1}_{0}(E)  +
\sum_{\gamma=1}^3  \bar{\delta}_{\beta \gamma} \, T_{\gamma}(E) 
\, G_{0}(E) U_{\gamma \alpha}(E)
\end{equation}
with $ \bar{\delta}_{\beta\alpha} = 1 - \delta_{\beta\alpha}$,
the free resolvent $G_0(E) = (E-H_0+i0)^{-1}$, and 
the free Hamiltonian for the relative motion $H_0$.
The two-particle transition matrices in three-particle space 
\begin{equation} 
T_{\gamma}(E) = v_{\gamma} + v_{\gamma} G_{0}(E) T_{\gamma}(E) 
\end{equation}
are obtained from the corresponding two-particle potentials
$v_{\gamma}$.
Odd-man-out notation is used to label pairs of particles and
two-cluster channels $\alpha$
with asymptotic states $| \Phi_\alpha \rangle$,
while on-shell matrix elements
 $\langle \Phi_\beta| U_{\beta \alpha}(E) | \Phi_\alpha \rangle$
yield  $\alpha \to \beta$ reaction amplitudes.
The integral equations (\ref{eq:Uba}) are solved in the momentum-space 
partial-wave framework, leading to well-converged results for 
 $\A{26}{Al}(d,p)\A{27}{Al}$ differential cross sections,
with the proton-nucleus Coulomb
force included via the screening and renormalization method.
Further technical details on the solution can be found in 
Refs.~\cite{deltuva:09b,deltuva:09d}.

\section{Results}

I study $\A{26}{Al}(d,p)\A{27}{Al}$ reactions using a number of 
realistic high-precision $np$ potentials: 
Argonne V18 (AV18) \cite{wiringa:95a},
charge-dependent Bonn (CD Bonn) \cite{machleidt:01a}, 
Reid93 \cite{stoks:94a},
chiral effective field theory ($\chi$EFT) potentials at 
next-to-next-to-next-to-next-to-leading
 order (N4LO) \cite{PhysRevLett.115.122301}
with regulators of 0.8 and 1.2 fm,
and the AV18 potential softened by the 
similarity renormalization group (SRG) transformation 
\cite{bogner:07b,bogner:07c} 
with the flow parameter $\lambda = 1.8\, \fm^{-1}$.
The deuteron $D$-state probability, characterizing
the strength of the tensor force and high-momentum components,
 acquires values in a broad range
from 2.53\% (SRG) to 5.76\% (AV18), even broader than in 
Ref.~\cite{PhysRevLett.117.162502}.
To compare with ADWA results of Ref.~\cite{PhysRevLett.117.162502},
I take the same NLOP  for 
$p$-$\A{26}{Al}$ and $n$-$\A{26}{Al}$, i.e.,
the parametrization of Giannini and Ricco \cite{giannini}
  without the spin-orbit part.
Binding potentials for $\A{27}{Al}$ are also chosen
as in Ref.~\cite{PhysRevLett.117.162502}, i.e.,
they have local Woods-Saxon form with the radius $r_0 = 1.25$ fm, 
diffuseness $a = 0.65$ fm, and spin-orbit strength $V_{so} = 6$ MeV,
whereas the strength of the central potential is adjusted to
the neutron separation energy $S_n$ for the given state.

\begin{figure}[!]
\begin{center}
\includegraphics[scale=0.66]{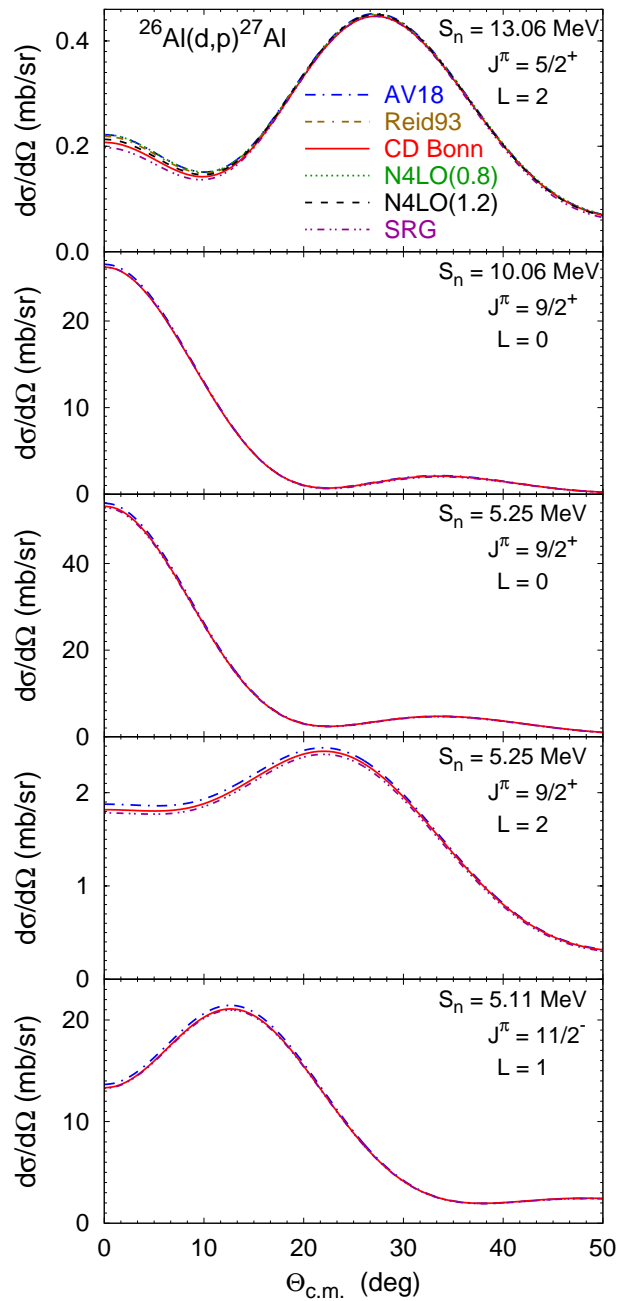}
\end{center}
\caption{\label{fig1}
Differential cross sections for 
$\A{26}{Al}(d,p)\A{27}{Al}$ reactions
at $E_d=12$ MeV as functions of the c.m. scattering angle.
The neutron separation energy $S_n$, spin and parity
of the final nucleus $J^\pi$, and the orbital angular momentum
transfer $L$ are indicated in each panel.
Predictions obtained with different realistic $np$ potentials are compared.}
\end{figure}

The results at the deuteron beam energy $E_d=12$ MeV
for  $\A{26}{Al}(d,p)\A{27}{Al}$ differential cross sections 
with ground and excited final states 
are presented in Fig.~\ref{fig1}. Both in the present work
and in Ref.~\cite{PhysRevLett.117.162502} the sensitivity
to the $np$ potential reaches its maximum at forward angles,
therefore the following comparison refers to the
 c.m. scattering angle $\Theta_\cm = 0$ deg.
The largest sensitivity to the $np$ potential using ADWA
in Ref.~\cite{PhysRevLett.117.162502}  was found for the 
transfer to the $\A{27}{Al}$ ground state with spin/parity $J^\pi = 5/2^+$
and $S_n = 13.06$ MeV. 
At $\Theta_\cm = 0$ deg the spread of ADWA  results
with realistic $np$ potentials in Ref.~\cite{PhysRevLett.117.162502}
is roughly a factor of 2.5, or about
$\pm 50$\% when measured from the central value.
In contrast, rigorous Faddeev calculations, as shown in the top panel of
Fig.~\ref{fig1}, exhibit much weaker sensitivity to the $np$ force model,
about $\pm 5$\% when measured from the respective central value.
 Furthermore, the magnitude of the differential cross section  predicted
using the Faddeev framework at $\Theta_\cm = 0$ deg 
 is lower by a factor of 2 (CD Bonn) to 4 (AV18) as compared to ADWA;
in the latter case even the shape is quite different.

Results for $(d,p)$ transfer reactions leading to
excited states of $\A{27}{Al}$ are also 
presented in Fig.~\ref{fig1}, this time using AV18, CD Bonn, and SRG
potentials only; predictions of other potentials lie between those of
AV18 and SRG and are therefore not shown.
In all cases the sensitivity to the $np$ force model is significantly
weaker than in  ADWA predictions of 
 Ref.~\cite{PhysRevLett.117.162502}. For example,
for the reaction leading to the 
$J^\pi = 9/2^+$, $S_n=5.25$ MeV state at $\Theta_\cm = 0$ deg 
the spread of Faddeev-type results, measured from their central values,  is
$\pm 1$\%  and $\pm 3$\% for the angular momentum transfer $L=0$
and $L=2$, respectively, while the spread of the corresponding
ADWA results is about $\pm 7$\%  and $\pm 25$\%.
Furthermore,
 the magnitude of Faddeev-type predictions for all excited states
shown in Fig.~\ref{fig1} is lower than ADWA
by a factor of roughly 1.5, while the shape is qualitatively similar.
Note that $np$ partial waves other than the deuteron wave
${}^3S_1-{}^3D_1$  contribute to the cross section up to
 10\%; those waves are neglected in ADWA \cite{PhysRevLett.117.162502}.

\begin{figure}[!]
\begin{center}
\includegraphics[scale=0.66]{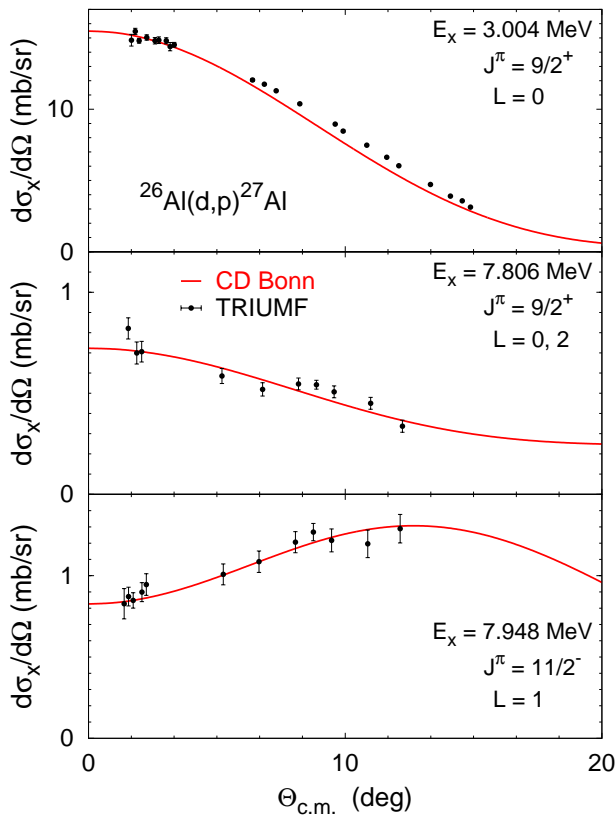}
\end{center}
\caption{\label{fig2}
Differential cross sections for 
$\A{26}{Al}(d,p)\A{27}{Al}$ reactions
at $E_d=12$ MeV as functions of the c.m. scattering angle.
The excitation energy $E_x$, spin and parity
of the final nucleus $J^\pi$, and the contributing 
orbital angular momentum
transfer $L$ are indicated in each panel.
CD Bonn predictions, rescaled by  factors $X(E_x,L)$ given in the text,
 are compared
with the experimental data from Ref.~\cite{PhysRevLett.115.062701}.}
\end{figure}

Although the main goal of the present work is the study of 
$(d,p)$ cross section sensitivity to the $np$ force model,
the comparison  of angular distributions with the experimental
data is also of some interest, even if the calculations
neglect the core excitation that is an important
dynamic ingredient. To account roughly for this shortcoming,
the theoretical predictions are rescaled by factors $X(E_x,L)$, 
such that the differential cross section ${d\sigma_{\mathrm{x}}}/{d\Omega}$
for the given  $\A{27}{Al}$ state 
with the excitation energy $E_x$  becomes
\begin{equation}
\frac{d\sigma_{\mathrm{x}}}{d\Omega} = \sum_L X(E_x,L) \, \frac{d\sigma(E_x,L)}{d\Omega}.
\end{equation}
I emphasize that $X(E_x,L)$ are not spectroscopic factors (SF),
but, given the low reaction energy $E_d = 12$ MeV,
 may be quite close to the SF as the rigorous study
including the core excitation indicates \cite{deltuva:17b}.
The  $X(E_x,L)$ values, obtained by adjusting the CD Bonn results
from Fig.~\ref{fig1} to the experimental data 
from Ref.~\cite{PhysRevLett.115.062701}, are 
$X(3.004,0) = 0.59$, $X(7.806,0) = 0.0105$, $X(7.806,2) = 0.09$, and
$X(7.948,1) = 0.062$, where the excitation energies are given in MeV.
The comparison is presented
in  Fig.~\ref{fig2}. One may conclude that theoretical
calculations describe the shape of the experimental data
\cite{PhysRevLett.115.062701} quite well.

\section{Discussion and conclusions}

The present study of $(d,p)$ reactions, performed in the
rigorous momentum-space three-body framework, indicates
low sensitivity to the $np$ potential model, and thereby
a failure of the ADWA with NLOP \cite{PhysRevLett.117.162502}. 
This may appear quite unexpected,
since a decent accuracy of the ADWA, 20\% or better 
\cite{nunes:11b,PhysRevC.95.064608},
was found when using local optical potentials. However, given
that no sensitivity to the $np$ model was observed for 
local optical potentials \cite{PhysRevC.95.024603}, 
much larger disagreement in the case of NLOP probably indicates the inadequacy
of ADWA specifically for treating NLOP. In fact, some inadequacy
can easily be seen when confronting the leading-order  NLOP effect
in ADWA \cite{PhysRevC.95.024603}
 to the rigorous Faddeev scattering framework \cite{faddeev:60a}.
This comparison is justified, since the leading-order yields
a good approximation of the complete ADWA  \cite{PhysRevC.95.024603}, 
and therefore shortcomings characteristic to the leading-order ADWA
 should be valid also in the case of the complete ADWA.
 The leading-order NLOP effect in ADWA is
 simulated replacing the energy-independent NLOP  in the initial deuteron 
channel by 
the equivalent local optical potential (ELOP) \cite{PhysRevC.95.024603}.
Energy-dependent ELOP has to be evaluated
 at the energy $E_{\rm loc} = E_d/2 + \Delta E$ where,
depending on the underlying $np$ potential, 
$\Delta E$ typically acquires values from $40 $ to 70 MeV.
E.g., $E_{\rm loc} \approx 50$ MeV (72 MeV) for CD Bonn (AV18)
models in the considered reactions at $E_d = 12$ MeV.
However, an inspection of the interaction terms
 on the r.h.s. of the Faddeev equation (\ref{eq:Uba}), i.e.,
 \begin{gather}  \label{eq:Eq}
\begin{split}
\int \langle \Phi_\beta | \bar{\delta}_{\beta \gamma}| p'_\gamma q_\gamma \rangle
d^3p'_\gamma 
\langle p'_\gamma | T_\gamma(E-q^2_\gamma/2M_\gamma)|p_\gamma \rangle \\
\frac{d^3p_\gamma d^3q_\gamma}{E+i0 - p^2_\gamma/2\mu_\gamma - q^2_\gamma/2M_\gamma}
\langle p_\gamma q_\gamma | U_{\gamma \alpha}(E) | \Phi_\alpha \rangle
\end{split}
\end{gather}
reveals that the energy $E-q^2_\gamma/2M_\gamma$
of any interacting two-particle subsystem
$\gamma$ with the intermediate relative momentum $p_\gamma$
is not fixed but depends on the respective 
spectator momentum $q_\gamma$ that is an integration variable,
formally ranging from zero to infinity;
$\mu_\gamma$ and $M_\gamma$ are the corresponding reduced masses.
The energy of the nucleon-nucleus subsystem therefore
formally acquires values from $-\infty$ to $E$,
i.e., its upper limit is 
$E_{NA}^{\mathrm{max}} = E_d\, A/(A+2) - 2.2245$ MeV,
amounting  to $E_{NA}^{\mathrm{max}} \approx 9$ MeV
in the present study.
Thus, $E_{\rm loc}$ of the order 50 MeV as required in the ADWA
appears to be a very unnatural value
from the Faddeev formalism point of view,
where the three-body amplitudes depend on fully
off-shell nucleon-nucleus transition operators 
$\langle p'_\gamma | T_\gamma(E-q^2_\gamma/2M_\gamma)|p_\gamma \rangle$
but with the relative two-body energy below $E_{NA}^{\mathrm{max}}$.
Note that there is no such a clear contradiction in the case
of ADWA with local potentials that are taken at the energy $E_d/2$.
Furthermore, the ELOP that is used in the leading-order ADWA is on-shell
equivalent to the original NLOP only at $E_{NA} = E_{\rm loc}$ but deviates
from it otherwise. As a consequence, ELOP and NLOP do not
provide an equivalent description of the initial scattering state
in $d+A$ collisions. Most importantly, 
the ability of the energy-independent NLOP to describe the
nucleon-nucleus subsystem over a broader energy range 
is lost when using  ELOP with fixed parameters \cite{giannini}. 

In conclusion, ADWA  appears to be inadequate for three-body 
 $(d,p)$ reactions with  NLOP, where low sensitivity to
the  $np$ interaction models is found by
rigorous Faddeev-type calculations. A similar conclusion
is drawn also by the alternative study using the approximate
continuum-discretized coupled-channel method
\cite{gomez:nlop}.
However, more significant sensitivity to nucleon-nucleon force models
can be expected in truly {\it ab initio} calculations without
any use of optical potentials, 
where even the threshold positions may depend on the force model.
Rigorous four-particle calculations of 
$\A{2}{H}(d,p)\A{3}{H}$ and $\A{2}{H}(d,n)\A{3}{He}$ reactions
\cite{deltuva:17a} provide an example for such dependence.

\begin{acknowledgments}
I thank E. Epelbaum for providing the codes for $\chi$EFT potentials.
I acknowledge the support  by the Alexander von Humboldt Foundation
under Grant No. LTU-1185721-HFST-E, and
the hospitality of the Ruhr-Universit\"at Bochum
where a part of this work was performed.
\end{acknowledgments}


\end{document}